\documentclass[conference,10pt]{IEEEtran}
\usepackage{amssymb}
\usepackage{amsmath}
\usepackage{graphicx}

\begin{document}

\title{Cooperative Multi-Cell Networks: Impact of Limited-Capacity Backhaul and
Inter-Users Links}

\author{\authorblockN{Shlomo Shamai (Shitz)\authorrefmark{1},
Oren Somekh\authorrefmark{2}, Osvaldo Simeone\authorrefmark{3},
Amichai Sanderovich\authorrefmark{1},} \authorblockN{Benjamin M.
Zaidel\authorrefmark{1}, and H. Vincent Poor\authorrefmark{2}}
\authorblockA{\authorrefmark{1}
Department of Electrical Engineering, Technion, Haifa 32000,
Israel} \authorblockA{\authorrefmark{2} Department of Electrical
Engineering, Princeton University, Princeton, NJ 08544, USA}
\authorblockA{\authorrefmark{3} CWCSPR, Department of Electrical
and Computer Engineering, NJIT, Newark, NJ 07102, USA}}

\maketitle

\begin{abstract}
Cooperative technology is expected to have a great impact on the
performance of cellular or, more generally, infrastructure
networks. Both multicell processing (cooperation among base
stations) and relaying (cooperation at the user level) are
currently being investigated. In this presentation, recent results
regarding the performance of multicell processing and user
cooperation under the assumption of limited-capacity inter-base
station and inter-user links, respectively, are reviewed. The
survey focuses on related results derived for non-fading uplink
and downlink channels of simple cellular system models. The
analytical treatment, facilitated by these simple setups, enhances
the insight into the limitations imposed by limited-capacity
constraints on the gains achievable by cooperative techniques.
\end{abstract}

\vspace{-0.1 cm}
\section{Introduction}
%\vspace{-0.1cm}

The performance limitations of conventional cellular wireless
networks in terms of throughput and coverage are by now well
recognized. This is due to extensive deployment of 2G and 3G
systems, and has stimulated the search for new approaches to
alleviate these drawbacks. A key technology that has been identified
to fulfill this goal is cooperation, to be employed at either the
base station (BS) or mobile station (MS) levels. As far as the BS
level is concerned, \textit{multi-cell processing} (MCP), sometimes
referred to also as a \emph{distributed antenna system}, prescribes
joint encoding/decoding of the signals transmitted/received at the
BSs through the exploitation of the high-capacity backbone
connecting the BSs (see
\cite{Shamai-Somekh-Zaidel-JWCC-2004}\cite{Somekh-Simeone-Barness-Haimovich-Shamai-BookChapt-07}
for recent surveys on MCP). Cooperation at the MS level in the
context of cellular networks has been studied under different names,
such as \textit{mesh}, \textit{hybrid} or \textit{multi-hop}
cellular networks, and is based on specific forms of relaying by the
MSs (see, e.g., \cite{pabst}).

BS and MS cooperative technologies are enabled by the presence,
respectively, of \textit{inter-BS (backbone) }and\textit{\
inter-MS links} that are not exploited by conventional cellular
systems for the purpose of encoding or decoding. These links can
be either wireless, orthogonal or not, thus possibly affecting the
interference or bandwidth budget of the network, or wired, hence
requiring additional deployment efforts.

Analysis of MCP (i.e., BS cooperation) has been so far mostly based
on the assumption that all the BSs in the network are connected to a
central processor via links of \textit{unlimited capacity}. In this
case, the set of BSs effectively acts as a multiantenna transmitter
(downlink) or receiver (uplink) with the caveat that the antennas
are geographically distributed over a large area. Since the
assumption of unlimited-capacity links to a central processor is
quite unrealistic for large networks, more recently, there have been
attempts to alleviate this condition by considering alternative
models. In \cite{somekh-zaidel-shamai} a model is studied where only
a subset of neighboring cells is connected to a central joint
processor. In \cite{atkas}\cite{shental} a topological constraint is
imposed where there exist unlimited capacity links only between
adjacent cells, and message passing techniques are employed in order
to perform joint decoding in the uplink. Finally, reference
\cite{sanderovich} focuses on the uplink and assumes that the links
between all the BSs and a central processor have limited capacity
(\textit{limited-capacity backhaul}). The reader is referred to
%\cite{marsch-ITW06}\ -\ \nocite{marsch-EW07}\cite{marsch-ICC07} for
\cite{marsch-EW07}\cite{marsch-ICC07} for another framework which
deals with practical aspects of limited capacity backhaul cellular
systems incorporating MCP.

Information-theoretic analysis of MS cooperation in cellular
networks is a more recent development. References include
\cite{simeone twc}\cite{somekh} where the uplink of a two-hop mesh
network is studied with \textit{amplify-and-forward} (AF)
cooperation (half-duplex and full-duplex, respectively) and
\cite{simeone-df} (half-duplex) \cite{simeone-allerton}
(full-duplex) where \textit{decode-and-forward} (DF)\ cooperation is
investigated (a thorough tutorial on cooperation techniques can be
found in \cite{kramer-marc-yates}).

\begin{figure}
\begin{center}
\includegraphics[scale=0.35]{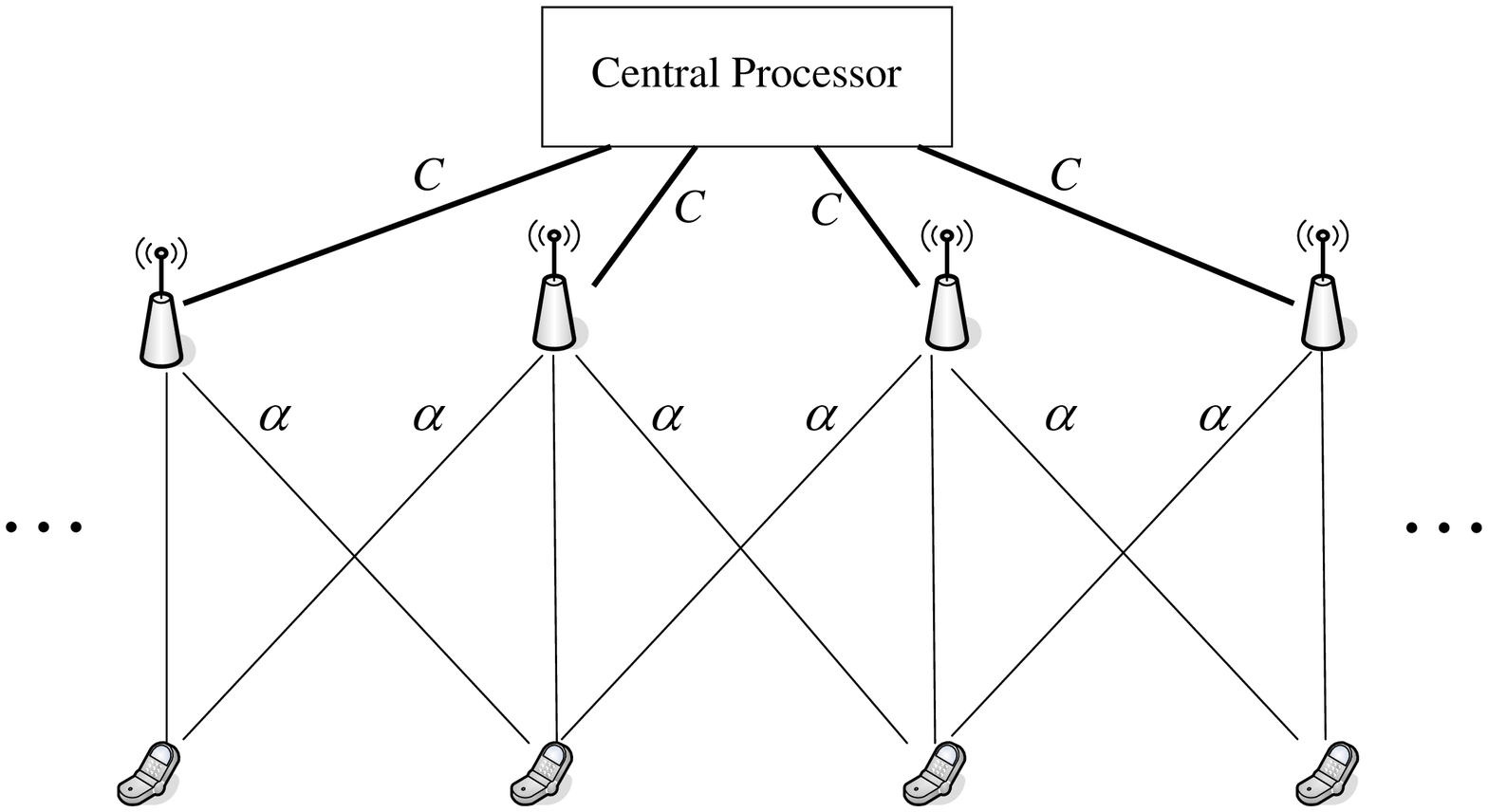}
\caption{Linear Wyner model with limited-capacity backhaul.}
\label{fig1}
\end{center}
\end{figure}

Most of the analysis on MCP is based on different variants of a
simple and analytically tractable model for a cellular system
proposed by Wyner \cite{Wyner-94} (henceforth, \textit{the Wyner
model}, see also \cite{Hanly-Whiting-Telc-1993}). Accordingly, the
cells are arranged in either an infinite linear array or in the
more familiar two-dimensional hexagonal pattern, and only
adjacent-cell interference is present and characterized by a
single gain parameter $\alpha\in(0,1]$ (see Fig. \ref{fig1}). In
some cases, we will also refer to a variation of the regular Wyner
model, called the \textit{``soft-handoff'' model}, where, assuming
a linear geometry, MSs are located at the border between two
successive cells and thus communicate only with the two
corresponding BSs. This model has been proposed in
\cite{Somekh-Zaidel-Shamai-IT-2005} (see also
\cite{Somekh-Zaidel-Shamai-CWIT-05}) and later adopted in a number
of works
\cite{Lifang-Goldsmith-Globecom06}\nocite{Lifang-Yoo-Goldsmith-Globecom06}
-\nocite{Jing-Tse-Hou-Soriaga-Smee-Padovani-ITA-2007}
\cite{Jing-Tse-Hou-Soriaga-Smee-Padovani-ISIT-2007}. With
simplicity and analytical tractability in mind, the Wyner model
provides perhaps the simplest framework for a cellular system that
still captures the essence of real-life phenomena such as
intercell interference and fading.

In this presentation, we focus on cellular systems abstracted
according to the Wyner model, and study the impact of
limited-capacity links on both MCP (\textit{limited-capacity
backhaul}) and MS cooperation (\textit{conferencing}) for the uplink
and the downlink. Moreover, we limit the scope to non-fading
Gaussian scenarios and to the case where only one user is active in
each cell at any given time (as for \textit{intra-cell TDMA}). It is
noted that intra-cell TDMA was proved to be optimal for non-fading
channels and the regular Wyner model in \cite{Wyner-94}, but no
claim of optimality is made here for the cases of limited-capacity
inter-BS and inter-MS of interest.

\vspace{-0.1cm}
\section{Limited-capacity backhaul}
\vspace{-0.1cm}

Analysis of MCP with limited-capacity backhaul can be carried out
according to different assumptions regarding the knowledge of
codebooks (or, more generally, encoding functions) at the BSs
(\textit{codebook information, CI}). In the sequel, we will treat
separately uplink and downlink channels.

\vspace{-0.1cm}
\subsection{Uplink Channel}
\vspace{-0.1cm}

In \cite{sanderovich}\cite{sanderovich-fullpaper}, the uplink of a
Wyner model with MCP and limited-capacity backhaul (see Fig.
\ref{fig1}) was studied in two scenarios: (\textit{i}) the BSs are
oblivious to the codebooks used by the MSs (no CI) so that
decoding is exclusively performed at the central processor;
(\textit{ii}) the BSs are aware of the codebooks used by the local
and the nearby MSs (cluster CI).

With oblivious BSs (case (\textit{i})), the cellular uplink
channel is equivalent to the setup of non-cooperative nomadic MSs
communicating with a central receiver via oblivious access points
with limited capacity links studied in
\cite{Amichai-fullpaper}\cite{Amichai-Conf_ISIT2006}. Focusing on
non-fading Gaussian channel and using the tools of
\cite{Amichai-fullpaper} combined with the inherent symmetry of
the Wyner model, an achievable per-cell sum-rate is given in
\cite{sanderovich}\cite{sanderovich-fullpaper} in the form of a
simple fixed-point equation. This rate presents an SNR penalty
with respect to the performance of the unlimited-capacity setup
\cite{Wyner-94}, but it coincides with the cut-set bound (taken
over the wireless and wired channels respectively) for high-SNR
and high backhaul capacity regimes. For the low-SNR regime,
\cite{sanderovich-fullpaper} shows that the fixed-point equation
characterizing the rate can be approximated to a closed-form
solution. Using this result, the low-SNR parameters of this
achievable rate, namely the minimum energy per-bit required for
reliable communication and respective low-SNR slope
\cite{Verdu-paper-low-snr-regime-02}, can be expressed as
functions of the low-SNR parameters of the unlimited setup and the
backhaul capacity $C$.

In the case of cluster CI (case (\textit{ii}) above), an
achievable rate is derived in
\cite{sanderovich}\cite{sanderovich-fullpaper} by allowing partial
local decoding at the base stations. According to this approach,
each MS splits its message and transmitted power into two parts:
one is intended to be decoded locally by the in-cell BS and
transmitted over the limited link to the central processor, while
the second part is processed according to the oblivious scheme and
is jointly decoded by the central processor. Assuming that each BS
is aware of the codebooks of the three MSs received by its antenna
(according to the limited-propagation property of the linear Wyner
model), it maximizes the local rate by selecting between
multiple-access and interference approaches \cite{ShamaiWyner97I}.
This local rate appears in the fixed-point equation mentioned
earlier since less bandwidth is now available for the oblivious
scheme applied for the second message part. Since the resulting
overall per-cell rate is non-concave, time-sharing is beneficial.
Nevertheless, numerical calculations reveal that a good strategy
is to do time-sharing between the two extremes: using decoding at
the BSs, with no decoding at the central processor, and doing
decoding only at the central processor, rather than using the
mixed approach. Since joint decoding is useful only when the BSs'
signals are correlated, the improvement in performance over the
oblivious scheme is increasing when the interference factor
$\alpha$ decreases, assuming that the limited capacity link does
not restrict the local rate. Interestingly, it is shown in
\cite{sanderovich}\cite{sanderovich-fullpaper} that incorporating
inter-cell time-sharing (ICTS) where the cells interfere each
other only a fraction of the time (see \cite{ShamaiWyner97I}) does
not provide significant improvement over the partial local
decoding scheme at hand.

\begin{figure}
\begin{center}
 \includegraphics[scale=0.5]{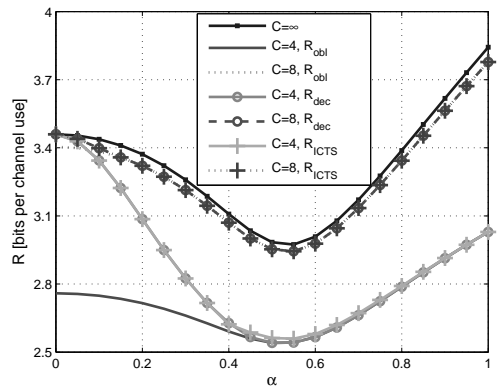}
 \caption{The uplink achievable rates $R_{obl}$, $R_{dec}$ and $R_{ICTS}$
 are plotted vs. the inter-cell interference $\alpha$ for backhaul capacity $C=4,8,\infty$, and user SNR $P=10$ [dB].}
 \label{fig:linear_asympt_comparison}
 \end{center}
\end{figure}

Achievable rates of the three schemes considered, are plotted in
Fig. \ref{fig:linear_asympt_comparison} for SNR, $P=10$ [dB], and
several backhaul capacity values, as functions of the inter-cell
interference factor $\alpha$. The gain of the cluster CI scheme
$R_{dec}$ over the oblivious scheme $R_{obl}$ when $C$ is low, is
prominent for low inter-cell interference. The cluster CI with
ICTS scheme $R_{ICTS}$ provides only a slight improvement over the
cluster CI scheme $R_{dec}$. As expected, for $\alpha=0$, the
rates achieved by the cluster CI and ICTS schemes are optimal,
when $C$ is larger than the respective unlimited setup rate (upper
bound), since no inter-cell interference is present and local
decoding with no ICTS is optimal.

Other settings such as the soft-handoff model and fading channels
are also considered in \cite{sanderovich-fullpaper} but are not
mentioned here for the sake of conciseness.

\vspace{-0.1cm}
\subsection{Downlink Channel}
\vspace{-0.1cm}

Turning to the downlink channel of the system in Fig. \ref{fig1},
MCP in the form of joint encoding in the framework of the
soft-handoff Wyner-like model is studied in
\cite{simeone-downlink} under the assumption of limited-capacity
backhaul. Similarly to \cite{sanderovich}
\cite{sanderovich-fullpaper}, three scenarios are considered that
present different tradeoffs between global processing at the
central unit and local processing at the BSs, with different
requirements in terms of CI at the BSs: \textit{(a)} local
encoding with CI limited to a subset of adjacent BSs (cluster CI);
\textit{(b)} mixed local and central encoding with only local CI;
\textit{(c)} central encoding with oblivious cells (no CI). Three
transmission strategies are proposed that provide achievable rates
for the considered scenarios.

Let us start with the case of cluster CI (case \textit{(a)}).
Exploiting the local interference structure of the soft-handoff
setup, shutting off one every $(J+2)$th BSs forms isolated
clusters of $(J+1)$ cells (see also \cite{lapidoth}). Each BS is
aware of the codebooks of its cluster's users, while the central
processor sends each cluster's messages to all its BSs via the
limited capacity links. Having the cluster's CI and messages, each
BS performs a form of \emph{dirty paper coding} (DPC) locally
(under individual equal per BS power constraint \cite{Yu-IT-2006})
and transmits its signal accordingly. In \cite{simeone-downlink}
two cluster DPC schemes are considered: 1) \emph{sequential}
encoding in which each BS invokes DPC to cancel the interfering
signal coming from its left neighboring BS; and 2) \emph{joint}
encoding in which each BS performs optimal joint DPC within the
cluster. It is shown in \cite{simeone-downlink} that the per-cell
rate of the joint encoding scheme, although it approaches the
cut-set bound when both the limited capacity $C$ and the cluster
size $(J+1)$ go to infinity while their ratio converges to some
finite constant, is in general lower than the rate of sequential
encoding for relatively small values of $C$.

In case \textit{(b)} of local CI, a scheme is proposed in
\cite{simeone-downlink} whereby each BS receives from the central
processor through the limited capacity link its local user's
message and a quantized version of the signal to be transmitted by
its left neighboring BS. By performing local DPC against the
quantized signal, where the inherent quantization error is treated
as an additional independent Gaussian noise, each BS is then able
to reduce the interfering signal (or cancel its quantized version)
coming from its left neighboring BS. The per-cell rate of this
scheme is given in \cite{simeone-downlink} as the unique solution
of a fixed-point equation. The closed-form expression derived in
\cite{simeone-downlink} reveals that the rate approaches the
cut-set bound only when the SNR is high and the interference level
is low.

With oblivious BSs (case \textit{(c)}), joint DPC under individual
equal power constraint is performed by the central processor,
which sends quantized versions of the transmitted signals to the
BSs via the limited-capacity links. Since the transmitted
quantization noise decreases the overall SNR seen by the MSs,
joint DPC is designed to meet lower SNR values and tighter power
constraints than those of the unlimited setup
\cite{Somekh-Zaidel-Shamai-IT-2005}. As expected, the resulting
per-cell rate is shown in \cite{simeone-downlink} to approach the
cut-set bound with increasing $C$. Moreover, also in the high-SNR
regime the scheme performs well achieving rates which are less
than 1 [bit/ channel use] below the cut-set bound.

The main conclusions of \cite{simeone-downlink} is that central
processing, even with oblivious BSs, is the preferred choice for
small-to-moderate SNRs or when the backhaul capacity $C$ is
allowed to increase with the SNR. On the other hand, for high SNR
values and fixed capacity $C$, a system with oblivious BSs is
limited by the quantization noise, and knowledge of the codebooks
at the BSs becomes the factor dominating the performance.
Therefore, in this regime, transmission schemes characterized by
local CI or cluster CI coupled with local processing outperforms
central processing with oblivious cells.

%\begin{figure}
%\begin{center}
%\includegraphics[scale=0.40]{vsCn.eps}
%\caption{The downlink achievable rates $R_{local}$ (cluster CI
%with local processing), $R_{mixed}$ (local CI with local and
%central processing), $R_{central}$ (no CI and central processing),
%and upper bound $R_{upper}$ (unlimited backhaul capacity) are
%plotted vs. the backhaul capacity $C$ for $P=10$ [dB] and
%$\alpha=1$.} \label{fig: Limited BH downlink}
%\end{center}
%\end{figure}

\begin{figure}
\begin{center}
\includegraphics[scale=0.40]{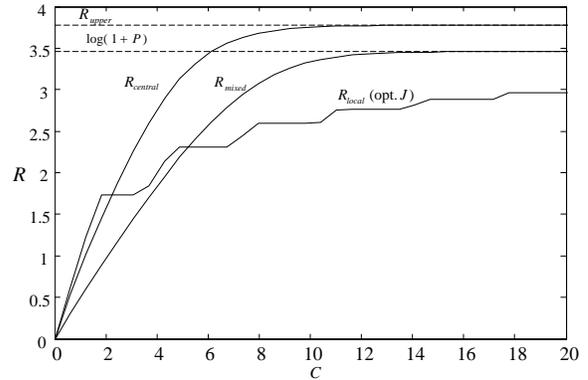}
\caption{The downlink achievable rates $R_{local}$, $R_{mixed}$,
$R_{central}$, and upper bound $R_{upper}$ are plotted vs. the
backhaul capacity $C$ for $P=10$ [dB] and $\alpha=1$.} \label{fig:
Limited BH downlink}
\end{center}
\end{figure}

Fig. \ref{fig: Limited BH downlink} shows the rates achievable by:
(a) local processing and cluster CI $R_{local}$ (with sequential
DPC and optimized $J$), (b) mixed (local and central) processing
and local CI $R_{mixed}$, and (c) central processing and no CI
$R_{central}$, versus the backhaul capacity $C$ for $P=10$ [dB]
and $\alpha=1$. It is noted that the optimal cluster-size $(J+1)$
is increasing with the capacity $C$ (not shown). It is seen that
if $C$ is large enough, and for relatively small to moderate
values of $P$, scheme 3, which performs central processing with
oblivious BSs $R_{central}$, is to be preferred.

\vspace{-0.1cm}
\section{Conferencing with limited-capacity inter-MS links}
\vspace{-0.1cm}

\begin{figure}
\begin{center}
\includegraphics[scale=0.35] {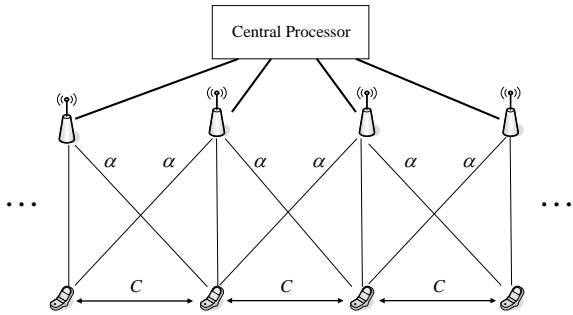} \caption{Linear Wyner model with
limited-capacity inter-MS conferencing.} \label{fig2}
\end{center}
\end{figure}

We now direct attention to a scenario where BSs perform multi-cell
processing (here, with unlimited-capacity backhaul), while the MSs
are allowed to cooperate over limited-capacity links (see Fig.
\ref{fig2}). These links should be considered as additional
spectral resources (orthogonal to the main uplink or downlink
channels) that are available for cooperation.

In modeling the interaction among MSs, the framework of
conferencing encoders for the uplink \cite{willems} (see also
\cite{maric1} for related scenarios) and decoders for the downlink
\cite{maric2}\nocite{servetto}\nocite{liang}\ -\ \cite{draper} is
followed. Moreover, as stated above, we focus on the non-fading
linear Wyner model with intra-cell TDMA so that conferencing
channels exist only between MSs belonging to adjacent cells
(\textit{inter-cell conferencing}) (reference
\cite{simeone-uplink} also considers intra-cell conferencing for
the uplink).

Starting with the uplink and multicell decoding (MCP), an upper
bound to the per-cell rate is obtained by considering a system
with perfect inter-MS cooperation whereby all the MSs are able to
exchange the local messages with all the other active MSs in the
network. The system at hand is thus equivalent to an
\emph{inter-symbol interference} (ISI) channel with channel state
information (CSI) at the transmitter (or equivalently an infinite
MIMO system with a Toeplitz channel matrix), for which a
stationary input with power spectral density obtained via standard
waterfilling is known to be optimal (in terms of sum-rate and
thus, by symmetry of the system at hand, in terms of per-cell
rate). An achievable rate can be derived by considering an
extension of the approach in \cite{willems} to multiple sources
(in the spirit of \cite{slepian-wolf}, Sec. VII)
\cite{simeone-uplink}. Specifically, rate splitting is performed
at each MS so that one message (the \textit{common} message) is
communicated to the $2J$ nearby MSs ($J$\ on either side) in $J$
rounds of conferencing on the limited-capacity links, so that in
the transmission phase cooperative transmission by $(2J+1)$ MSs
can take place for every common message. Based on the observation
that a stationary input is asymptotically optimal, cooperative
transmission can be designed so as to implement an equivalent
linear pre-filtering of the transmitted signal, which in the limit
of $C\rightarrow\infty$ (and $J\rightarrow\infty)$ allows the
upper bound discussed above to be attained via appropriate design
of the filter at hand \cite{simeone-uplink}. It is worth
mentioning that the number of conferencing rounds $J$, restricts
the time-span of the impulse response of the pre-coding filter.

\begin{figure}
\begin{center}
\includegraphics[scale=0.40]{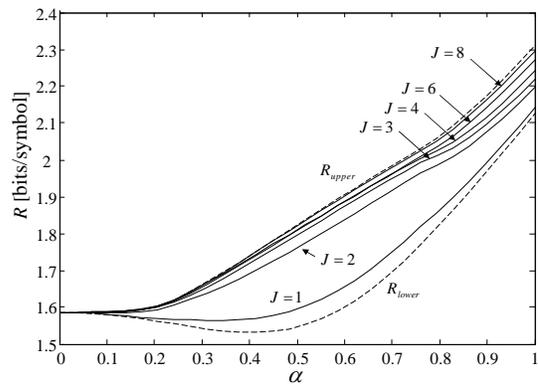}
\caption{The achievable rate $R$ (with inter-cell conferencing and
intra-cell TDMA), $R_{upper}$, and $R_{lower}$, are plotted vs.
the inter-cell interference factor $\alpha$ for $P=3$ [dB], $C =
10$, and several values of $J$.} \label{fig: Conf uplink}
\end{center}
\end{figure}

In Fig. \ref{fig: Conf uplink} the achievable rate $R$ is plotted
vs. the inter-cell gain $\alpha $ along with the lower bound
$R_{lower}$ (no cooperation) and upper bound $R_{upper}$ for
$C=10$ and $P=3$ [dB]. Note that very relevant performance gains
can be obtained by increasing the number of conference rounds,
especially from $J=1$ to $J=2$. Moreover, having sufficient large
conferencing capacity $C$ and number of conference rounds $J$
(with $C/J\geq R_{upper})$ enables for the upper bound to be
approached. It is worth mentioning that increasing $J$ is always
beneficial to obtain a better approximation of the waterfilling
strategy. However, due to the limited conferencing capacity $C$,
it is not necessarily advantageous in terms of the achievable rate
(not shown).

When considering the downlink, the upper bound on the rate
achieved with perfect cooperation is again the rate of an ISI\
channel with CSI, for which a waterfilling power spectral density
of the input is optimal. Since we are interested here in the
processing to be performed at the receiving end, that is at the
MSs, it is relevant to point out that optimal decoding in case of
perfect cooperation can be implemented by means of a Minimum Mean
Square Error - Decision Feedback Equalizer (MMSE-DFE)
\cite{guess}. Moreover, the same per-cell rate (recall the
symmetry argument) can also be achieved by moving the DFE\ part of
the equalizer to the transmitter via DPC \cite{zamir} \cite{yu1}.
An achievable, and asymptotically ($C\rightarrow\infty$) optimal,
rate can then be obtained by having the central encoder perform
DPC to cancel the post-cursor interference, while the MSs exchange
quantized versions of the received signals on the conferencing
channels in order to enable MMSE filtering on a subset of channel
outputs. Analysis of this scenario involves the problem of source
encoding with side information on the conferencing channels, and
is currently under study.

\vspace{-0.1cm}
\section{Concluding Remarks}
\vspace{-0.1cm}

Three decades after their introduction, the information theoretic
understanding of cellular systems is far from being complete. In
its full generality, it touches upon the most basic information
theoretic models, not yet fully understood. Those comprise
combinations of multiple-access, broadcast, interference and relay
MIMO frequency selective fading channels, as well as fundamental
network information theoretic aspects. With the new setup at hand,
which incorporates limited capacity links among MSs, and between
BSs and central processor, it is expected that information theory
will continue to play a central role in assessing the ultimate
potential and limitations of cellular networks as well as in
providing fundamental insights into the architecture and operation
of future systems.

\vspace{-0.1cm}
\section*{Acknowledgment}
\vspace{-0.1cm}

The research was supported in part by a Marie Curie Outgoing
International Fellowship and the NEWCOM++ network of excellence
both within the 6th European Community Framework Programme, by the
U.S. National Science Foundation under Grants CNS-06-25637 and
CNS-06-26611, and also by the REMON Consortium.

\end{document}